\documentclass[aps,prd,reprint,groupedaddress,superscriptaddress]{revtex4-2}

\usepackage[dvipdfmx]{graphicx}
\usepackage{color}

\usepackage{amsmath,amsfonts,amsthm,bm,ulem}

\newcommand{\TT}[2]{\Lambda_{#1#2}}

\newcommand\MO{}

\newcommand\AH{}

\begin{document}

%\preprint{}

\title{Quark-model search for compact ${c}\bar{c}uds$ pentaquark states}
%\title{Progress Report: Quark model for a compact \textit{{udsc}$\bar{c}$} \textbf pentaquark state}

\author{Emiko Hiyama}
\email[hiyama@riken.jp]
\homepage
%\thanks{}
\affiliation{Department of Physics, Tohoku University,      Sendai~
          980-8578, 
          Japan}
   \affiliation{Few-body Systems in Physics Laboratory, RIKEN Nishina Center, Wako,
         Saitama 351-0198, Japan}
\affiliation{
Institute for Theoretical Physics, Heidelberg University, Philosophenweg 16,
      69120 Heidelberg,
          Germany}
          \author{Atsushi Hosaka}
\email[hosaka@rcnp.osaka-u.ac.jp]
          \homepage
\affiliation{Research Center for Nuclear Physics (RCNP), Osaka University,  
      Ibaraki, Osaka~
          {567-0047}, 
          Japan}
\affiliation{Advanced Science Research Center, Japan Atomic Energy Agency (JAEA) Tokai 319-1195, Japan}
\author{Makoto Oka}
 \email[makoto.oka@riken.jp]
 \homepage
   \affiliation{Few-body Systems in Physics Laboratory, RIKEN Nishina Center, Wako,
         Saitama 351-0198, Japan}
\affiliation{Department of Physics, Tohoku University,      Sendai~
          980-8578, 
          Japan}
\affiliation{Advanced Science Research Center, Japan Atomic Energy Agency (JAEA) Tokai 319-1195, Japan}
\author{Georg Wolschin}
        \email[wolschin@uni-heidelberg.de]
        \homepage
 \affiliation{Few-body Systems in Physics Laboratory, RIKEN Nishina Center, Wako,
         Saitama 351-0198, Japan}  
         \affiliation{
Institute for Theoretical Physics, Heidelberg University, Philosophenweg 16,
      69120 Heidelberg,
          Germany}   

\date{\today}

\begin{abstract}
    A potential quark model is used to search for a $P_{c\bar{c}s}^0=(c\bar{c}uds)^0$, $J^P=1/2^-$ pentaquark state that has recently been observed experimentally by the LHCb collaboration at 4338.2 MeV, with a width of 7.0 MeV and high statistical significance $>15\sigma$. 
Our model Hamiltonian reproduces the masses of the low-lying charmed and strange hadrons. We use the Gaussian expansion method {to solve the} five-body Schr\"odinger equation. 
Employing the real scaling method {including}
 the relevant meson-baryon thresholds explicitly, sharp resonances are distinguished from the meson-baryon
scattering states.
We incorporate new color states of the color-octet meson and baryon configurations as well as the color-singlet configurations for the five-quark states.
We find no $ J^P=1/2^-$ resonance close to the observed state, and also none in the $ J^P=3/2^-$ state. This increases the likelihood that $P^0_{c\bar{c}s}$ is a $\Xi_c\bar{D}$ hadronic molecule rather than a compact state. 
 \end{abstract}

%\keywords{}

\maketitle

\section{Introduction}
\label{intro}

\AH{Ever since the quark model \cite{gell64,peter65,zweig64} has been formulated, 
quests for multiquarks beyond conventional mesons and baryons have been continuing.  
A breakthrough came in 2003 by the two reports of evidences of the  $\Theta^+$ pentaquark~\cite{nak03} and 
the $X(3872)$ tetraquark~\cite{belle03}.  
The initial observation of $\Theta^+$ did not persist due to null results with better statistics \cite{hicks12,pra24}, 
while further efforts in the analysis by the LEPS2 group at SPring-8 are still ongoing~ \cite{nak25}.  
In contrast,  evidences have been accumulated for the $X(3872)$ due to the high statistics data from  
LHCb collaboration~\cite{LHCb:2011zzp}.
By now the existence of the peak structure almost at the value of the $D \bar D^{*0}$ threshold
has been confirmed.  
The discovery of the $X(3872)$ is now followed by the series of observations of non-standard hadrons, called $X, Y, Z$ mesons
with hidden charm~\cite{Hosaka:2016pey}, open charm tetraquark $T_{cc}$~\cite{lhcb20}, and also $P_{c\bar{c}}^+=(c\bar{c}uud)^+$ pentaquarks~\cite{lhcb15,lhcb19}.}

\MO{These new ``exotic'' states have several common features, (1) they contain heavy quark(s), 
(2) they often lie close to a two-hadron threshold and (3) many of them look like molecular bound states of two heavy hadrons.
Therefore it is interesting and important to study whether ``genuine'' compact tetra- or pentaquark states exist or not.}

The LHCb collaboration has recently observed a strong resonance corresponding to a $P_{c\bar{c}s}^0=(c\bar{c}uds)^0$, $J^P=1/2^-$ pentaquark candidate in $B^-\rightarrow J/\psi\Lambda\bar{p}$ decays \cite{lhcb23} with high statistical significance $>15\sigma$. Earlier results with indications for possible $P_{c\bar{c}s}^0, J^P=1/2^-,3/2^-$ states at higher invariant masses in the decay $\Xi_b^-\rightarrow J/\psi\Lambda K^-$ \cite{lhcb21} could add to these findings, but so far have less statistical significance of $3.1\sigma$.

In the current project, we investigate the possible existence of compact $c\bar{c}uds$, pentaquark states of $ J^P=1/2^-,3/2^-$, 
by using a nonrelativistic quark model. 
Constituent quark models have proven to be useful in many investigations that could subsequently be refined in more rigorous treatments of quantum chromodynamics such as lattice calculations and sum rules.
\AH{Our strategy is to solve the Schr\"odinger equation for five-body systems  as precisely as possible by the  Gaussian expansion method (GEM). 
In contrast to ordinary baryons of three quarks the pentaquark baryons may fall apart into a three-quark baryon and a quark-antiquark meson.  
Hence the coupling of the ``confined" pentaquarks to the scattering meson-baryon system is crucial for the existence of the stable of resonant pentaquarks. }

The present work builds upon previous investigations, where we have studied 
resonances in $c\bar{c}uud$  \cite{emiko18}  and $c\bar{c}sss$  \cite{emiko19} systems, as candidates of compact pentaquark states. However, for $c\bar{c}uud$ a calculated $1/2^-$ resonance at 4690 MeV is more than 230 MeV too high in mass compared to the current experimental LHCb values of 4312 and 4457 MeV \cite{lhcb19}, and a second $3/2^-$ resonance at 4920 MeV is 480 MeV above the measured value of 4440 MeV. 
Hence, the above resonances from the quark-model calculations are unlikely to represent the observed states, which instead appear to be 
$\Sigma_c^+\bar{D}^{*0}$ molecular states with $J^P=1/2^- (E_B\simeq 3\,\text{MeV})$ 
and $3/2^- (E_B\simeq 20\, \text{MeV})$.
Similarly, we have found several resonances in the $c \bar c sss$ systems with narrow widths, which, however, are waiting for 
confirmation by experiments.

In the present work, we turn to the $P_{c\bar{c}s}=(c\bar{c}uds)^0$, $J^P=1/2^-$ state that was found by the LHCb collaboration in the $J/\psi\Lambda$ system produced in pp collisions at 7, 8 and 13 TeV. 
It emerges as an outstanding invariant-mass peak 
just 0.83 MeV above the $\Xi_c^+D^-$  
threshold within the experimental error bars. 
The estimated mass and width are $M=4338.2\pm 0.7\pm 0.4$ MeV and $\Gamma=7.0\pm 1.2\pm1.3$ MeV, respectively,  with 
the statistical significance is above $15\sigma$ \cite{lhcb23}. 
In analogy to $P_{c\bar{c}}(4312)^+$, it could be a hadronic molecule (e.g. virtual $\Xi_c^{+}{D}^-$ state with $E_B\simeq -0.83$.

Earlier LHCb data on the $P_{c\bar{c}s}(4459)$ pentaquark in the decay $\Xi_b^-\rightarrow J/\psi\Lambda K^-$  with less statistical significance of 3.1$\sigma$ could be described by a double-peak structure with 13 MeV mass splitting, $P_{c\bar{c}s}(4455)$,  and $P_{c\bar{c}s}(4468)$ and respective decay widths of $\Gamma=7.5\pm 9.7$  MeV and $5.2\pm 5.3$ MeV~\cite{lhcb21,Karliner:2022erb}. 
This  could be analogous to the $P_c(4440), \  P_c(4457)$ with possible $J^{P}=1/2^-$ and $3/2^-$. 
The Belle collaboration has recently confirmed the existence of the 
$P_{c\bar{c}s}(4459)$ pentaquark resonance with a local significance of $3.3\, \sigma$ 
using the first observations of $\Upsilon(1S, 2S)$ inclusive decays to $J/\psi\Lambda$ 
in $e^+e^-$ collisions \cite{belle25}, but they do not yet find evidence for a double peak, and give 
only upper limits for the LHCb peak at 4338 MeV.

It  is our aim to investigate whether these states could appear as resonances in a full five-body calculation in the non-relativistic quark model.  
Technically, we follow the lines of Ref.\,\cite{emiko19}. 
We expect that the results will clarify whether the signals observed by LHCb is consistent with a compact $c\bar{c}uds$ pentaquark state, or closer to a molecular state.  
Different from the previous calculation \cite{emiko19} for $c\bar{c}sss$, 
we now include new color-octet flavor-singlet meson-baryon configurations, which are favored by 
the color-spin interaction among the light quarks, and thus could be important to form the pentaquark resonance \cite{Irie:2017qai,Takeuchi:2016ejt,oka20,Giachino:2022pws}.

{In Sec.\,\ref{sec2}, the model Hamiltonian and the choices of the variational wave function for the 5-quark system 
are reviewed. 
The real scaling method to identify resonances from the meson-baryon scattering states is explained also in Sec.\,\ref{sec2}.
In Sec.\,\ref{sec3}, the results for $ J^P=1/2^-,3/2^-$ states are displayed and discussed. The summary and outlook is given in Sec.\,\ref{sec4}.}

%%%%%%%%%%%%%%%% 
\section{The model}
\label{sec2}

The method of solving the five-body Schr\"odinger equation with a realistic quark-model potential\,\cite{emiko03,emiko06,rich17} has been described in detail in Refs.\,\cite{emiko18,emiko19}.  
We consider not only bound-state solutions, but also states in the continuum. The coupling to the scattering states is of crucial importance, because states that appear when the coupling is absent may disappear when scattering states are included. 

To identify a resonance, the technique of real scaling along the fall-apart coordinate is used, starting from a variational function that includes many types of clustering, and accounting for all possible open channels such as $\Lambda_c^++D_s^{-}$ at 4255 MeV, $\Xi_c^0+\bar{D}^{0}$ at 4319 MeV, etc.

\subsection{Hamiltonian}

We solve the five-body Sch\"odinger equation 
with the Hamiltonian in the nonrelativistic quark model, given by
\begin{align}
\label{hamiltonian}
H=&\sum_i^5\left(m_i+\frac{{\bf{p}}_i^2}{2m_i}\right)-T_{\rm cm}\\\nonumber
&-\frac{3}{16}\sum_{i<j=1}^5\sum_{a=1}^8(\lambda^a_i\lambda^a_j)V_{ij}({\bf{r}}_{ij}) \, .
\end{align}
Here $m_i$ and ${\bf{p}}_i$ are the respective mass and momentum of the $i$th quark, $T_{\rm cm}$ is the kinetic energy of the center-of-mass motion, and $\lambda_i^a (a=1,...8)$ are the color SU(3) matrices for the $i$th quark with color index $a$. 
\MO{Note that $\lambda^a$ is replaced
by $(-\lambda^{a*})$ for the anti-quark ($i=5$).} 
The interaction potential is adapted from Semay and Silvestre-Brac \cite{se94,si96}
\begin{align}
V_{ij}({\bf{r}}_{ij})=&-\frac{\kappa}{r}+\lambda r^p-\Lambda\nonumber\\
&+\frac{2\pi\kappa '}{3m_im_j}\frac{\exp(-r^2/r_0^2)}{\pi^{3/2}r_0^3}{\bm{\sigma}_i}\cdot{\bm{\sigma}_j}\,.
\label{Vpotential}
\end{align}
It consists of the Coulomb potential, the confining part, a color-electric constant term, and the color-magnetic spin-spin 
{(color-spin)} interaction term with Pauli's matrices ${\bm{\sigma}_i}, {\bm{\sigma}_j}$. The cutoff parameter $r_0$ depends on the quark masses,
\begin{align}
r_0(m_i,m_j)=A\left(\frac{2m_im_j}{m_i+m_j}\right)^{-B}\,.
\end{align}

In the present calculation, we choose the AP1 potential with $p=2/3$ \cite{se94,si96}. 
The model parameters $\kappa, \kappa ', \lambda, \Lambda, A$ and $B$ are given in {Table}\,\ref{parameters}. 
With these parameters, the meson and baryon masses relevant for the thresholds in the present $c\bar{c}uds$ calculations 
are reasonably well reproduced.

\MO{The last term of Eq.\,(\ref{Vpotential}) is the color-spin term, 
which represents the exchange of a color-magnetic gluon between quarks.
It plays an important role in the hadron spectroscopy, giving
the mass splittings of the pseudoscalar ($0^-$) and vector ($1^-$) mesons 
as well as the spin $1/2$ and $3/2$ baryons.
As the magnetic coupling of the gluon to the quark is inversely proportional to the quark mass,
the effect of the color-spin term is suppressed for heavy quarks.}
Then, in the heavy quark limit ($m_c\to\infty$), the spin of the heavy quark decouples from the Hamiltonian and is conserved independently
from the light quark spins.
This is called the heavy-quark spin symmetry (HQSS).
\MO{We will see in the next subsection that the color-spin potential favors a new color-octet configuration 
$(uds)_{\bf 8}$ in the case of the $P_{c\bar c s}$ pentaquark.\,\cite{Irie:2017qai,Takeuchi:2016ejt,oka20,Giachino:2022pws}. }

\begin{table}[htp]
%\begin{center}
\caption{The AP1 ($p=2/3$) potential parameters \cite{se94,si96} and the calculated and experimental masses of the relevant strange/charm hadrons.}\label{parameters}
\begin{tabular}{|c|c||c|c|c|c|c}
\hline
\multicolumn{2}{|c||}{Parameters of}& \multicolumn{3}{c|}{Hadrons masses (MeV/c$^2$)}\\
%\hline
\multicolumn{2}{|c||}{the AP1 potential}&hadron& calculated & experimental\\
\hline
$m_{u,d}$ (MeV)&277&$\eta_c$&2984&2984  \\
$m_s$ (MeV)&553&$J/\psi$&3103&3097\\
$m_c$ (MeV)&1819&$D$&1881&1865\\
&&$D^*$&2033&2007\\
&&$D_s$&1940&1968\\
&&$D_s^*$&2097&2112\\
\hline
$p$&2/3&$\Lambda$&1151&1116\\
$\kappa$&0.4242&$\Sigma$&1240&1193\\
$\kappa'$&1.8025&$\Lambda_c$&2286& 2287\\
$\lambda$ (GeV$^{1+p}$)&0.3898&$\Sigma_c$&2465&2454\\
$\Lambda$ (MeV)&11313&$\Xi_c$&2498&2469\\
$A$ (GeV$^{B-1}$)&1.5296&$\Xi'_c$&2606&2579\\
$B$&0.3263&$\Xi^*_c$&2679&2646\\
\hline
\end{tabular}
%\end{center}
\end{table}

\subsection{Jacobi coordinate systems}

\begin{figure}[htbp]
 \includegraphics[width=5.5cm]{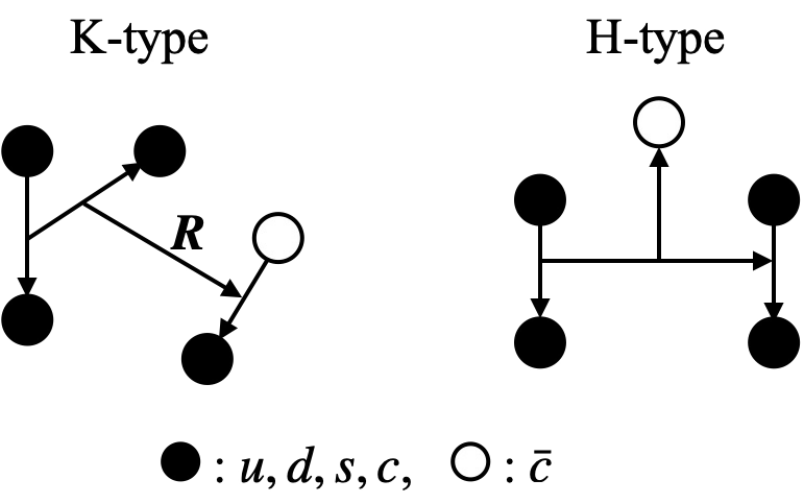}
\caption{
	\label{fig1}
Types of the Jacobi coordinate systems for the GEM calculation. The K-type configurations $(3q)+(q\bar q)$ have two choices of color assignments as explained in the text. For the H-type we consider only the $\overline{\bf 3}$ diquarks. $R$ is the relative coordinate between the baryon and the meson, which is scaled by a factor $\alpha$ 
in the real scaling method.}
\end{figure}

When we apply the Gaussian Expansion Method (GEM) to the pentaquark,  
multiple Jacobi coordinate systems (called configurations) are introduced 
and the variational wave function is given as a superposition of the configurations, denoted by C,
\begin{align}
\Psi  =&\sum_C a_C \Psi^{(C)} \, .
\label{wave-function-C}
\end{align}
We consider two types of the Jacobi coordinates (Fig.\,\ref{fig1}), {\it i.e.,}
the ``K'' type as a product of a 3-quark cluster and a quark-antiquark cluster, 3+2, and 
the ``H'' type as a product of 2-quark+ 2-quark + antiquark clusters. 
For each type, several different wave functions are prepared with
various possible spin, flavor and color quantum numbers for each cluster.
It should be noted that the terms in Eq.\,(\ref{wave-function-C}) are, in general, 
not orthogonal to each other.

For the ``color-singlet'' K(1) type, 
we consider a product of color-singlet $(uds)_{\bf 1}$ and $(c\bar{c})_{\bf 1}$, 
corresponding to $\Lambda\eta_c$ and $\Lambda J/\psi$ configurations. The other combinations include
products of color-singlet $cud/csu/csd$ and $\bar{c}u/\bar{c}d/\bar{c}s$ configurations, 
corresponding to the $\Lambda_c$, $\Sigma_c^{(*)}$, $\Xi_c^{(*)}$ baryons, and the $D^{(*)}$, $D_s^{(*)}$ mesons.
Altogether, we take four different configurations, 
$(uds) (c\bar c), (udc) (s\bar c), (ucs) (d\bar c), (cds) (u\bar c)$ in the  K(1) type. 
 For each configuration, we have several spin combinations for the baryon and meson as well.

In this study, as a new color configuration, we introduce the ``color-octet'' K(8) type configurations,
given by products of a color-octet
3-quark cluster and a color-octet quark-antiquark cluster.
For instance, we incorporate the combination of $(uds)_{\bf 8}$ and $(c\bar{c})_{\bf 8}$, 
where both the $(uds)$ or $(c\bar c)$ form color-octet {\bf 8} states,
and they are combined into the totally color-singlet 5-quark system.
We have four configurations for K(8).
The K(8) configurations differ only by the color combination and the other parts, {\it i.e.,} spin, orbital, isospin wave functions
are identical to the K(1) configurations.

The color wave function for K(8) is explicitly given by
\begin{align}
|(q\bar q)({\bf 8},a) \rangle =& \frac{1}{\sqrt{2}} \sum_{k,\ell} \lambda^a_{k,m} |\bar q (\overline{\bf 3},k)\rangle |q; ({\bf 3},m)\rangle\\
|(qq)(\overline{\bf 3},k)\rangle=& \frac{1}{\sqrt{2}}\sum_{mn} \epsilon_{kmn} |q; ({\bf 3},m)\rangle|q; ({\bf 3},n)\rangle\\
|(qqq)({\bf 8},a)\rangle=&|(qq)_{\overline{\bf 3}} q; ({\bf 8},a)\rangle  \nonumber\\
=& \frac{1}{\sqrt{2}} \sum_{k,m} \lambda^a_{k,m} |(qq) (\overline{\bf 3},k)\rangle |q; ({\bf 3},m)\rangle\\
|(qqq)_{\bf 8}, (q\bar q)_{\bf 8}; {\bf 1}\rangle=&\frac{1}{\sqrt{8}} \sum_a |(qqq)({\bf 8},a)\rangle |(q\bar q)({\bf 8},a)\rangle,
\end{align}
\MO{where $m,n =(1,2,3)$ are the indices for the color triplet representation of the quark and 
$k=(1,2,3)$ is for the anti-triplet representation of the antiquark.}
In Appendix \ref{appB}, we present some matrix elements of the color operator, $\sum_a \lambda^a_i \lambda^a_j$, 
which {appear} in the Hamiltonian.

The importance of the color-octet configurations comes from the color-spin term of the Hamiltonian, the last term in Eq.\,(\ref{Vpotential}).
To demonstrate the importance of the color-spin interaction,
we consider the heavy quark limit for simplicity, and focus on the color-spin state of the light quarks, $uds$.
Further assuming that in a compact state, the orbital wave function is common to all three quarks,
we use the following formula to evaluate the matrix element of the color-spin operator $\Delta$\,
(See Appendix \ref{appA} for the derivation and also see Refs.\,\cite{Jaffe:1976yi,Oka:2023hdc}), 
\begin{align}
& \Delta= - \sum_{i<j} \sum_a (\lambda_i^a\lambda_j^a) (\bm\sigma_i\cdot\bm\sigma_j)\\
& \langle\Delta\rangle = n(n-10)+\frac{4}{3}S_n(S_n+1) +2\,C_2[C_n] +4\,C_2[F_n],
\end{align}
where $n$ is the number of quarks ($n=3$ in this case), 
$S_n$ is the total spin, and $C_n$ and $F_n$ represent the color and flavor irreducible representations of SU(3).
$C_2$ is the SU(3) quadratic Casimir defined in Eq.\,(\ref{C2}).
From this formula one sees that the color-spin term of the Hamiltonian favors the flavor anti-symmetric state.
In the case of $(uds)$, the flavor singlet state is totally antisymmetric. It, however, happens that 
the Pauli principle does not allow to take the color-singlet and flavor-singlet state simultaneously 
with the totally symmetric orbital configuration. Thus the color-octet state becomes the best choice now.
Namely, $\langle\Delta\rangle= -14$ for $(qqq)_{\bf 8_c, \bf 1_f}$, 
compared with $\langle\Delta\rangle= -8$ for $(qqq)_{\bf 1_c, \bf 8_f}$.
The coefficient of $\Delta$ in the Hamiltonian can be estimated from the baryon spectrum, $M(\text{decuplet})-M(\text{octet})\sim 300$ MeV $\sim 16\langle\Delta\rangle$.
Then, the difference in the energy between the color-octet and color-singlet states is roughly given by 110 MeV.

The H-type configurations are chosen so that diquark $qq$ clusters belong to the color $\overline{\bf 3}$ representation,
and including the antiquark, three color $\overline{\bf 3}$ states are combined into the total color singlet state as
\begin{align}
&|(qq)_{\overline{\bf 3}}, (qq)_{\overline{\bf 3}}, \bar q; {\bf 1}\rangle\nonumber\\
=&\frac{1}{\sqrt{6}} \sum_{i,j,k} \epsilon_{ijk}
|(qq) (\overline{\bf 3},i)\rangle|(qq) (\overline{\bf 3},j)\rangle |\bar q(\overline{\bf 3},k)\rangle\,,
\end{align}
\MO{where $(i,j,k)$ runs over the antisymmetric combinations of $(1,2,3)$}.
Another possible choice of the color of $qq$ cluster is {\bf 6}, but we only choose $\overline{\bf 3}$ states in this calculation, 
{because the $\overline{\bf 3}$ diquark is favored by the color-spin interaction 
(by typically about 200 MeV $\sim (M(\Sigma^{(*)}_c)-M(\Lambda_c))$.}
We take two H-type configurations, $(ud)(sc)\bar c$ and $(us)(dc) \bar c$.

Both the K(8)-type and H-type configurations are ``confined''. 
Namely, all the radial coordinates are not allowed to go beyond the confinement scale.
In contrast, the color-singlet K(1)-type configurations contain two color-singlet clusters, 
which can be separated arbitrarily. 
Then, for each configuration $C$, there is one radial coordinate, denoted by $R_C$,
which is the relative distance between the color-singlet clusters, {\it i.e.,} a baryon and a meson.
Therefore the radial wave function for $R_C$ can represent scattering states of two hadrons.
Later, we use $R_C$ for the real scaling analysis.

Finally, the total 5-body wave function is arranged to have a definite angular momentum $(J,M)$.
In total, the five-body wave function is represented by a superposition given as
\begin{align}
\Psi_{JM}=&\sum_C A^{(C)} \xi_1^{C}\sum_\gamma B_\gamma^{C}
\left[\chi_S^C\phi_L^C\right]_{JM}\,,
\label{GEM}
\end{align}
where $A^{(C)}$ and $B_\gamma^{C}$ are variational constants that indicate the contribution of each configuration.
The $\xi_1^{C},\chi_S^C$, and $\phi_L^C$ are the color, spin and spatial wave functions, respectively, 
for total color singlet, spin $S$ and orbital angular momentum $L$. The index $\gamma$ is introduced to label
all these quantum numbers collectively.\,\cite{emiko19}.
In all cases, the total wave function must be properly antisymmetrized.

GEM superposes several (of order 10) Gaussian functions with different range parameters 
for each radial coordinate of each configuration in Eq.(\ref{GEM}). Thus they form non-orthogonal bases states for
the variational method. Namely, the matrix elements of the Hamiltonian for the bases states are calculated
and the Hamiltonian is diagonalized by solving the non-orthogonal eigenvalue equation.
Both positive and negative parity states can be computed in this scheme, although we focus on negative parity states with orbital angular momentum $L=0$ and total spin-parity $J^P=1/2^-,3/2^-$,
to be compared with the experimentally observed states.
About \MO{$43,000$ ($26,000$)} basis functions are used to diagonalize the Hamiltonian
 for $J^P=1/2^{-}( 3/2^{-})$. 
 
\subsection{Real Scaling Method}

In the pentaquark system, it is inevitable to consider the meson-baryon thresholds. Namely, the five quark system 
can dissociate into a meson and a baryon once the threshold is open (fall-apart decay). This is different from the ordinary quark model
calculation, where for instance a three quark state will decay into a meson and baryon only through processes 
of creating a $q\bar q$ pair. Thus, if we ``neglect'' the pair creation, the threshold will not open. This is not the case
for the tetra- or {pentaquark} systems.

For the pentaquark system, if there is a compact state that couple{s} weakly
to the meson-baryon scattering states, it may appear as a resonance. 
Here in GEM, we identify the resonance states above the meson-baryon threshold by using the real scaling method.

Since the system is computed within a finite volume in GEM, the resulting eigenvalues are discrete even for baryon-meson scattering solutions. 
As in the earlier works \cite{emiko06,emiko18}, the real-scaling (stabilization) method \cite{sim81} is used to distinguish true resonances from discretized scattering states. 
The basis functions are scaled along $R^{C}$ for the K(1)-type configurations
by multiplying every range parameter with a factor $\alpha$, $R^C \rightarrow \alpha R^C$, with $1\le\alpha\le 2$. 
The continuum states will then fall off towards their respective thresholds.
\MO{In contrast, a compact resonant state will not be affected by the boundary at large distance,
as the wave function does not extend beyond the range of the color confinement.}

It is noted that molecular states like $\Xi_c\bar{D}$ or $\Xi_c^{'}\bar{D}$ are expected to be bound by attractive meson exchange forces as discussed in the introduction. They are not included in the present calculation, because in our model there is no residual interaction once two color-singlet hadrons are separated in the K(1)-type configurations. 
A part of the short-range interactions between the baryon and meson is, however, included in the present calculation through quark exchanges -- even between the color-singlet hadrons.
It is one of the purposes of the present work to reveal whether 
such short-range interaction could provide sufficient attractive interaction, or not.

\section{Results}

\label{sec3}

\subsection{{Results of the confined H+K(8) configurations}} \label{sec3a}

In order to examine the roles of different configurations and see how the real scaling method works,
we first present the calculation only with the confined configurations, {\it i.e.,} the H and K(8) types.
This ``approximation'' corresponds to the standard quark model calculation, where the coupling to 
meson-baryon scattering channels are ignored and only the compact discrete states are presented.

\begin{figure}[htbp]
\includegraphics[width=7cm]{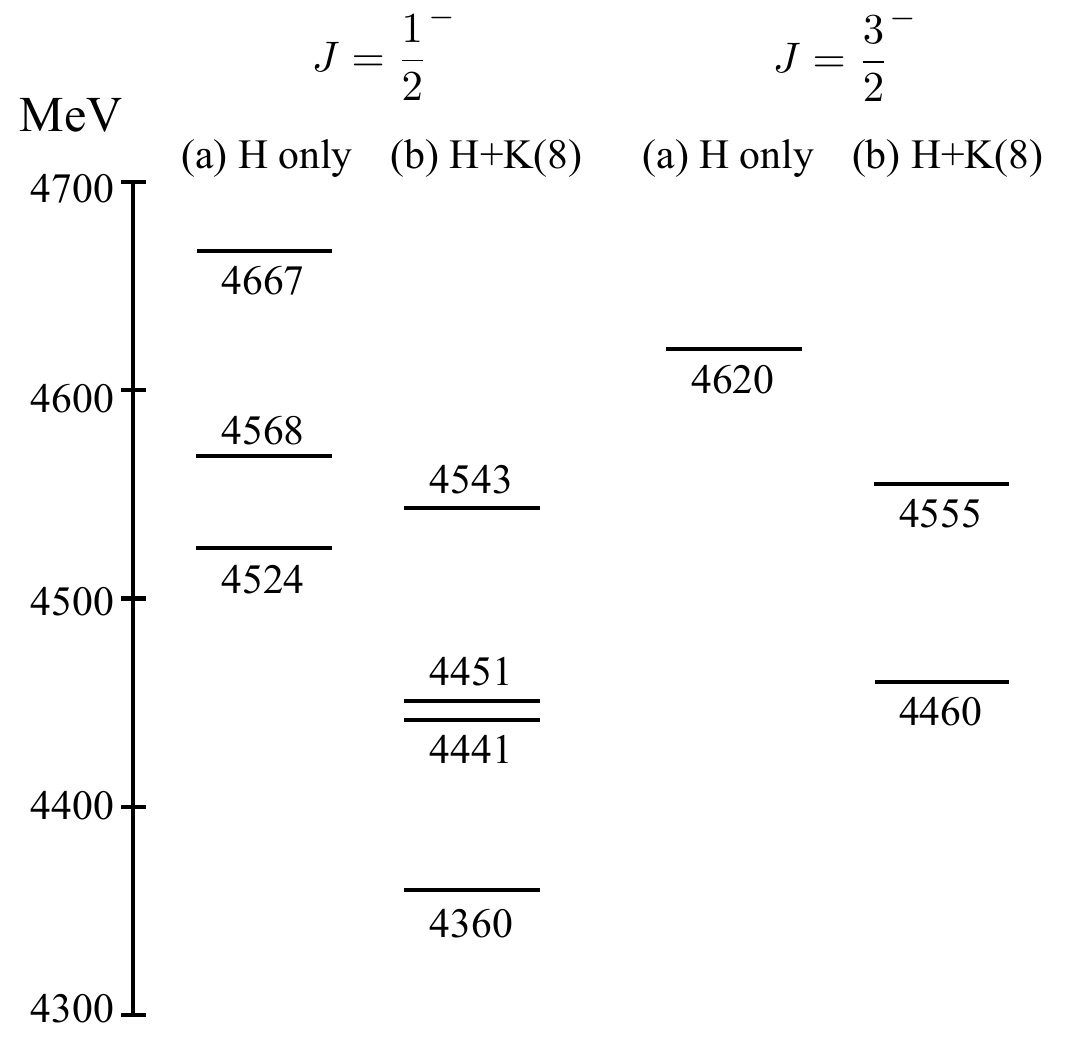}
\caption{Results of the calculation with the confined H and K(8) configurations for $J^P=1/2^-$ and $3/2^-$ states. 
}\label{fig2}
\end{figure}

\begin{figure*}[htbp]
\begin{center}
\includegraphics[width=12cm]{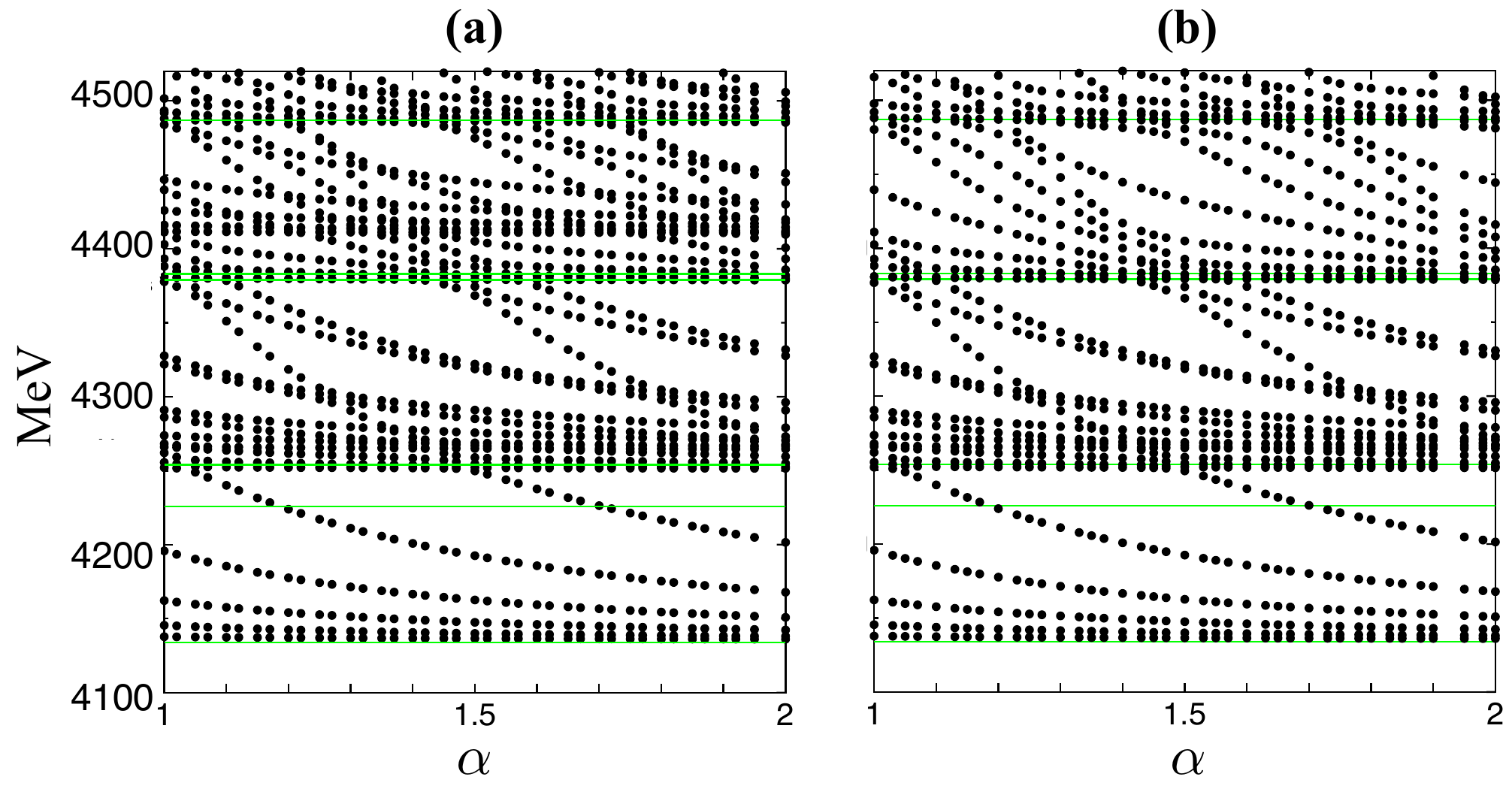}
\end{center}
\caption{Real scaling results for $J^P=1/2^-$. for the (a) H+K(1) and (b) H+K(8)+K(1)
configurations. 
The horizontal lines indicate the model values of the thresholds, 
$\Lambda\eta_c$, $\Lambda_cD_s$, $\Lambda J/\psi$, $\Xi_c D$,
$\Lambda_c D_s^*$ and $\Xi_c' D$ from bottom to top.}\label{fig3}
\end{figure*}

\begin{figure*}[htbp]
\includegraphics[width=12cm]{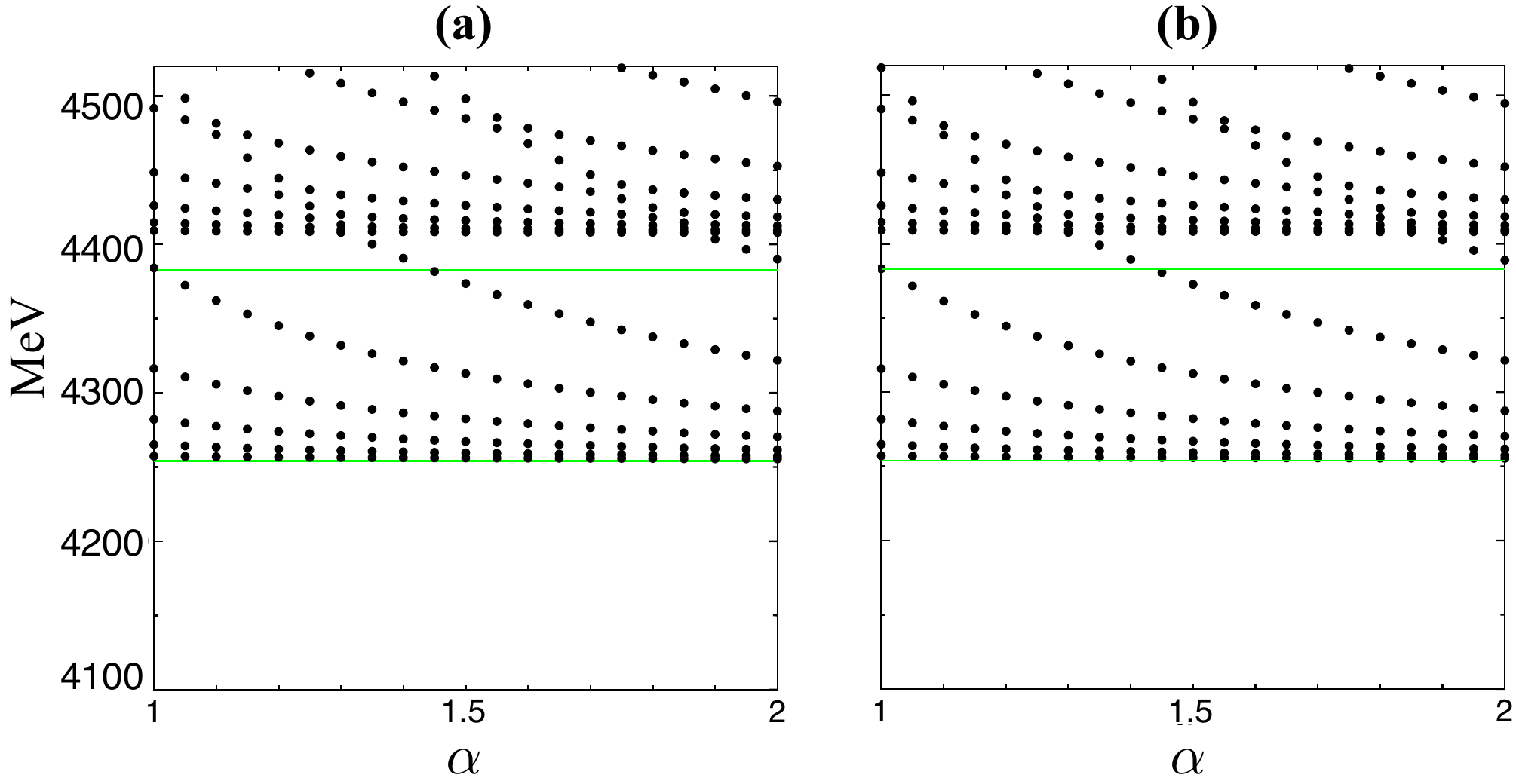}
\caption{Real scaling results for $J^P=3/2^-$ for the (a) H+K(1) and (b) H+K(8)+K(1)
configurations. 
The horizontal lines indicate the model values of the thresholds of $\Lambda J/\psi$ (lower) and
$\Lambda_c D_s^*$ (higher).}\label{fig4}
\end{figure*}

Fig.\,{\ref{fig2}} shows the ``discrete'' energy levels of 
the $P_{c\bar{c}s}$ with the spin-parity $J^P=1/2^-\,\text{and}~3/2^-$ from the GEM calculation only with the confined configurations.
We compare the results obtained (a) only with the H configurations, and (b) with the H and K(8) configurations.
We find no bound state, {\it i.e.,} all the eigenvalues are above the lowest meson-baryon hidden charm thresholds, $\Lambda+\eta_c$ at $4135$ MeV and  $\Lambda+J/\psi$ at $4254$ MeV. 
The lowest $J^P=1/2^-$ state in the H+K(8) calculation appears at 4360 MeV, followed by two closely lying states at 4441 MeV and 4451 MeV.
They lie just around the observed $P_{c\bar{c}s}(4455)/P_{c\bar{c}s}(4468)$, 
while it is higher by about 100 MeV than the $P_{c\bar{c}s}(4338)$ state.

The H+K(8) energy levels are lower than those from the H-only calculation by about 80-120 MeV.
It is consistent with the above simple estimation of the energy gain from the color-spin interaction. 
This confirms the significance of the color-octet configuration \cite{Irie:2017qai,Takeuchi:2016ejt,oka20,Giachino:2022pws}.

One sees that the $J^P=3/2^-$ states are higher than the $J^P=1/2^-$ states, as is expected from the spin-dependent part
of the Hamiltonian. 
It is also noted that some of the $J=1/2$ and $3/2$ states are paired into the heavy-quark spin-doublet,
which comes from the heavy-quark spin symmetry.

\begin{figure}[htbp]
 \includegraphics[width=8.5cm]{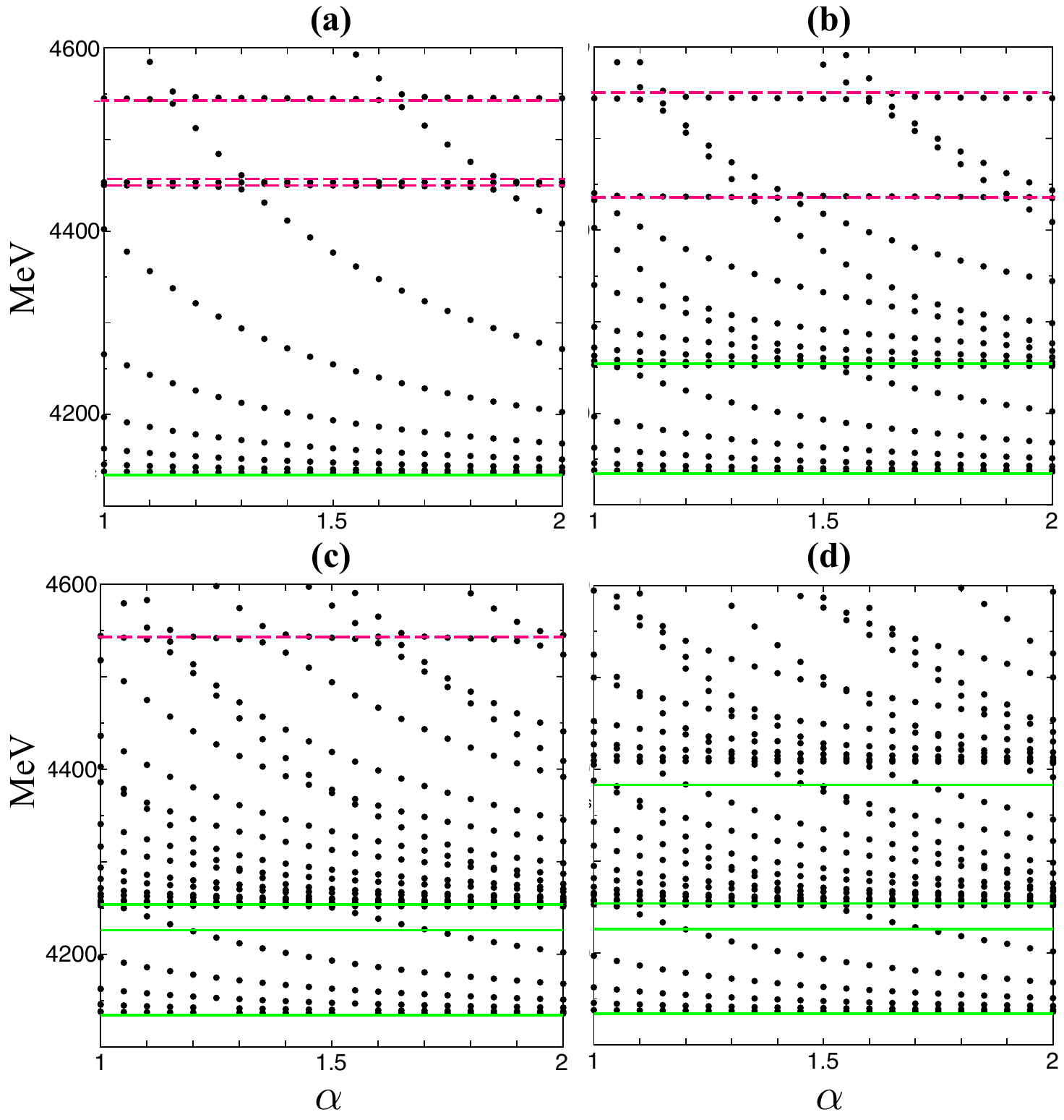}
\caption{Real scaling results for $J^P=1/2^-$ with selected open channels. 
(a) H +K(8) +K(1, $\Lambda\eta_c$), (b) H +K(8) +K(1, $\Lambda\eta_c+ \Lambda_cD_s$),
(c) H +K(8) +K(1, $\Lambda\eta_c+\Lambda_cD_s+\Lambda J/\psi$), 
(d) H +K(8) +K(1, $\Lambda\eta_c+\Lambda_cD_s+\Lambda J/\psi+ \Lambda_c D_s^*$).
The red-dashed lines indicate the ``resonance-like'' states and the green horizontal lines represent the thresholds}\label{fig5}
\end{figure}

\begin{figure}[htbp]
\includegraphics[width=8.5cm]{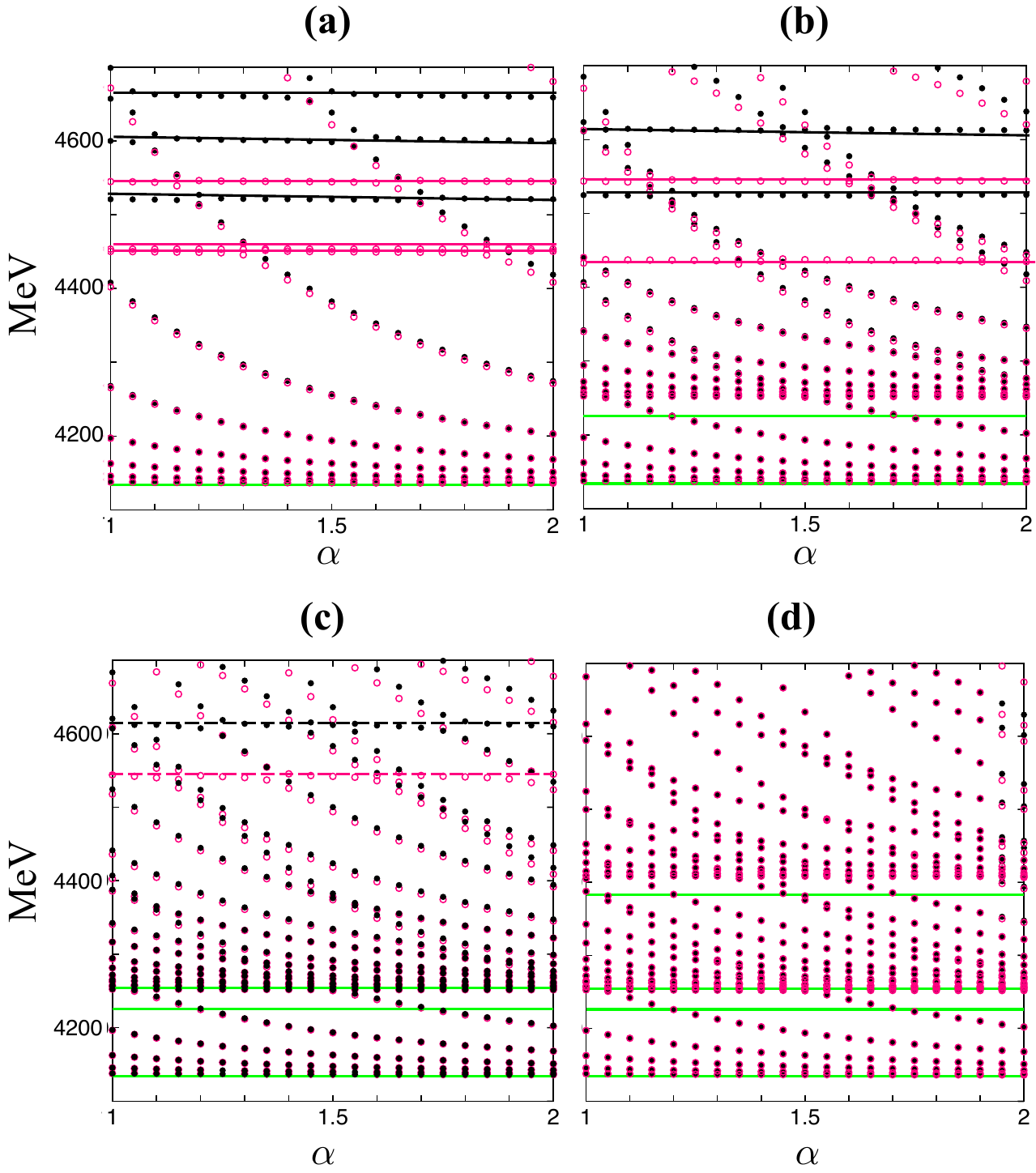}
\caption{
	\label{fig6}
Comparison between the real scaling calculations of the H+K(1) (black points) and H+K(8)+K(1) (red circles) configurations,
(a) with only K(1, $\Lambda\eta_c$)),
(b) $\Lambda_cD_s$ added, (c) $\Lambda J/\psi$ added and (d) $\Lambda_c D_s^*$ added.}
\end{figure}

\subsection{{Results including K(1) configurations}}\label{sec3b}

We have seen some discrete $P_{c\bar{c}s}$ states around the 4.4 GeV region in the H+K(8) calculation.
However, this calculation does not yet take into account couplings to meson-baryon scattering states, 
while multiple thresholds are already open in the relevant energy region. 
In order to complete our analysis, we need to include the K(1) configurations
in the GEM calculation and perform the real scaling method analysis.

Fig.\,\ref{fig3} and Fig.\,\ref{fig4} show the stabilization plots as functions of the scaling parameter $\alpha$ 
for the $c\bar{c}uds$ system in $J^P=1/2^-$ and $J^P=3/2^-$,
respectively.
As the GEM restricts the variational wave function within a finite range, all the eigenvalues are real and discrete.
In order to identify sharp resonances, we vary the scaling parameter $\alpha$ from 1 to 2, and plot the low-lying eigenvalues. 
One sees that the eigenvalues corresponding to scattering states descent 
towards the thresholds, indicated by the horizontal lines) as 
$\alpha$ increases.
On the other hand, a bound and/or resonance state should remain at the binding/resonance energy regardless of the value of the scaling parameter. 
Note that the horizontal lines are the calculated meson-baryon threshold energies. 
For some higher thresholds, 
they do not completely agree with the accumulation of the real scaling points. 
\MO{This is caused by the reduced number of the basis states in the full 5-body calculation, while the qualitative results and conclusions
are not affected.}

In Figs.\,\ref{fig3} and \ref{fig4}, we cannot find a stable state (corresponding to a sharp resonance), in the energy region, 
$4100 \lesssim \text{E}\lesssim 4500$  MeV,
which contains the region where the $P_{c\bar{c}s}$ states were observed. 
All the eigenvalues move towards the thresholds, thus indicating that the discrete states seen in the H+K(8) calculation melt away into the scattering states in the complete five-body calculation.

It is informative to study how the discrete ``quark model states'' in Fig.\,\ref{fig2} are mixed with the meson-baryon scattering states without forming sharp resonances. 
In the real scaling method, one can include only selected scattering channels,
and study which scattering channel couples to a discrete state one by one.
Fig.\,\ref{fig5} shows the real scaling results
for the $J^P=1/2^-$ state including (a) only the $\Lambda\eta_c$ continuum states, 
(b) the $\Lambda\eta_c+ \Lambda_cD_s$,
(c) the $\Lambda\eta_c+\Lambda_cD_s+\Lambda J/\psi$ and 
(d) the $\Lambda\eta_c+\Lambda_cD_s+\Lambda J/\psi+\Lambda_c D_s^*$
continuum states. 
One sees in Fig.\,\ref{fig5}(a) that the lowest discrete state at 4360 MeV disappears by the coupling of $\Lambda\eta_c$, 
while there remain three sharp resonance-like states at higher energies 
indicated by the horizontal lines.
When (b) the $\Lambda_c D_s$ channel is explicitly
taken into account, the state at 4441 MeV is melted away. 
Similarly, the higher states (4451 MeV and 4543 MeV) disappear when
(c) $\Lambda J/\psi$ and (d) $\Lambda_c D_s^*$ are included, respectively. 
These behaviors indicate that the ``discrete'' states seen in the H+K(8) calculation are
dominated by the low-lying baryon-meson scattering states.

It is also interesting to see the comparison between the H+K(1) and H+K(8)+K(1) calculations. 
In Fig. \ref{fig2}, one sees 
a large energy gain by adding the K(8) configuration. 
The physical interpretation of this gain is that the color-spin interaction
brings strong attraction for the color-octet flavor-singlet state. 
It is, however, not seen at the end for the full calculation
with the K(1) scattering states.

In Fig.\,\ref{fig6}, 
we compare the results for the H+K(1, selected) and H+K(8)+K(1, selected) calculations 
for $J^P=(1/2)^-$ states.
We find that the stable horizontal lines are significantly affected by the inclusion of
the K(8) configurations, as is consistent with the discrete levels shown in Fig. \ref{fig2}.
However, the $\Lambda-\eta_c$ scattering states coming from the K(1, $\Lambda\eta_c$) configurations are almost identical between
the H+K(1, $\Lambda\eta_c$) and H+K(8)+K(1, $\Lambda\eta_c$) calculations.
Similar results are seen in Fig. \ref{fig6} (b), (c) and (d), when we further add the $\Lambda_cD$, $\Lambda J/\psi$ and
$\Lambda_c D_s^*$ configurations successively.
This observation implies that the eigenstates obtained in the real scaling calculation are dominated by the scattering states, such as
the low lying $\Lambda\eta_c$, $\Lambda_cD_s$, and $\Lambda J/\psi$, and are not affected by the compact multi-quark configurations.
Then the discrete states from the confined (H or H+K(8)) calculation disappear when the K(1) scattering states are coupled 
to them.

\section{Conclusions and Outlook}
\label{sec4}

We have presented solutions of the five-body Schr\"odinger equation for compact $P_{c\bar{c}s}^0=(c\bar{c}uds)^0$, $J^P=(1/2^-,3/2^-)$ states in the energy region between 4100 MeV and 4500 MeV. Pentaquark states had been found experimentally by the LHCb collaboration at 4338 MeV ($J^P=1/2^-$) with high statistical significance $>15\sigma$, and at 4455  MeV ($J^P=1/2^-$) as well as 4468 MeV ($J^P=3/2^-$) with low significance $<3.1\sigma$.  Using a quark model Hamiltonian that reproduces the masses of the low-lying charmed and strange hadrons, we employ the Gaussian expansion method for an accurate solution of Schr\"odinger's equation. When including the relevant baryon-meson thresholds explicitly, the real scaling method allows us to distinguish sharp resonances from meson-baryon scattering states.

 We have considered two types of Jacobi coordinates: a product of a three-quark cluster plus a quark-antiquark cluster, and a product of two-quark plus two-quark plus antiquark cluster. For each type, wave functions with various possible spin, flavor and color quantum numbers have been considered. We include four color-singlet configurations,
with several spin combinations for the baryon and meson. 

As a new ingredient, we consider color-octet configurations,
given by products of a color-octet three-quark cluster and a color-octet quark-antiquark cluster. It turns out that this consideration of color-octet excitations is significant and produces an energy gain from the color-spin interaction: The color-octet configuration lowers the pentaquark energy.  However, the effect is not strong enough to keep the discrete states as resonances.
In the real scaling calculation, we find all the discrete ``quark model'' states 
melt away into the scattering states. At the end, we have found no resonance in the $J^P=(1/2^-,3/2^-)$ channels in the energy region of interest.

In contrast, compact five-quark states with relatively narrow widths had previously been found using  quark-model calculations in the $(c\bar{c}sss)^0$ system.
This discrepancy between $c\bar{c}sss$ and $c\bar{c}uds$ could be a consequence of the stronger binding in the system with three strange quarks,
and would provide indirect evidence that not only the $P_{c\bar{c}}^+$ states found by LHCb, but also the newly detected $P_{c\bar{c}s}^0$ state(s) are most likely molecular meson-baryon configurations bound by meson exchange forces. From a recent model for hidden-charm pentaquarks with strangeness descibed as $\Lambda_c^{('*)} \bar{D}_s^{(*)}$ and  $\Xi_c^{('*)} \bar{D}^{(*)}$ molecules coupled to a five-quark core, the experimental LHCb values for the three measured $P_{c\bar{c}s}^0$ states could indeed be reproduced, albeit with adapted potential parameters \cite{gia23}. Many more results for molecular configurations are documented in Ref.\,\cite{rpp24}.

In our present calculation, part of the short-range interactions between the baryon and meson are included  through quark exchanges. However, ``real" molecular states like $\Xi_c\bar{D}$ or $\Xi_c^{'}\bar{D}$ that are bound by attractive meson exchange forces are not included, because in our model there is no residual interaction once two distinguishable color-singlet hadrons are formed.

In order to realize truly compact pentaquarks, it may be necessary to resort to systems with even stronger binding such as $P_{b\bar{b}c}$. An example is $(b\bar{b}usc)^+$, which could, however, also be a $\Xi_b^0B_c^+$ hadronic molecule, or  $(b\bar{b}dsc)^0$, corresponding to $\Xi_b^-B_c^+$. These strongly bound systems may possibly turn out to be compact pentaquarks with $M\simeq 12$ GeV in future investigations. Their experimental detection in the $\Upsilon\, \Xi_c$ or other channels will, however, be very challenging.

\bigbreak
\begin{acknowledgments}
\MO{We acknowledge Sachiko Takeuchi for fruitful discussions. 
This work is supported in part by JSPS Grants-in-Aid for Transformative Research Areas 
(Quantum Matter Science in the Universe Opened Up by Precise Numerical Calculations), 
JP25A203 and JP25H01267 (E.H. and M.O.).
It is also supported by JSPS Grants-in-Aid for Scientific Research JP23K03427 (M.O), JP24K07050 (A.H.) and 
by JST ERATO Grant Number JPMJER2304 (E.H.). 
G.W. is supported by the `Expanding Internationality' program of Heidelberg University, ExU 11.2.1.31,
that provided funding for E.H.'s stays in 2024 and 2025 at the Institute for Theoretical Physics, Heidelberg. 
G.W. is thankful for financial support during his 2025 stay at RIKEN.
The numerical calculation was carried out by the use of the supercomputer ``Genkai'' at Kyushu University.}
\end{acknowledgments}

\appendix

\newcommand{\vecsig}{\bm\sigma}

\section{Eigenvalues of the color-spin operator}\label{appA}

We calculate the eigenvalues of the color-spin operator,
\begin{eqnarray}
&& \Delta =-\sum_{i<j} (\vecsig_i\cdot\vecsig_j) (\lambda_i\cdot\lambda_j) \qquad \qquad\qquad \qquad
\end{eqnarray}
for general $n$-quark systems with totally symmetric orbital wave function. 
The eigenstates of $\Delta$ can be labeled by the total color $C_n$, total spin $S_n$ and total flavor $F_n$.
Denoting the spin, flavor and color exchange operators of the $i$ and $j$ quarks as
$P_{ij}^{s}$, $P_{ij}^{f}$ and $P_{ij}^{c}$, respectively,
we obtain
\begin{align}
&  (\vecsig_i\cdot\vecsig_j) = 2P_{ij}^{s} -1 \\
& \langle \sum_{i<j} P_{ij}^{s} \rangle = \frac{n(n-4)}{4} + S_n(S_n+1),\\
&  (\lambda_i\cdot\lambda_j) = 2P_{ij}^{c} -\frac{2}{3}\\
& \langle \sum_{i<j} P_{ij}^{c} \rangle = \frac{n(n-9)}{6} + C_2[C_n].
%& \langle \sum_{i<j} P_{ij}^{c} P_{ij}^{s}\rangle = \frac{n(n-36)}{12} +C^6_2[C_n,S_n] , \label{a2}
\end{align}
Here $C_2$ is the SU(3) quadratic Casimir
%$C_2 =\langle \sum_a T^a T^a \rangle $ with
%$T^a = \sum_i \lambda^a_i/2$, where $\lambda^a_i$ the color Gell-Mann matrix of the $i$-th quark.
%It is 
given by
\begin{eqnarray}
&& C_2[(p,q)] = \frac{p^2+pq+q^2}{3} +p+q
\label{C2}
\end{eqnarray}
for the irreducible representation $(p,q)$ of SU(3).

For the totally antisymmetrized $n$-quark system, (note that we consider only the totally symmetric orbital states)
we find for the flavor $F_n$ representation as
\begin{align}
& \langle \sum_{i<j} P_{ij}^{c} P_{ij}^{s}\rangle  
=- \langle \sum_{i<j} P_{ij}^{f}\rangle
= - \frac{n(n-9)}{6} - C_2[F_n].
\end{align}
Combining all the above, we obtain a general formula for an antisymmetrized $n$-quark system
\begin{align}
& \langle \Delta\rangle = -\frac{2}{3}\frac{n(n-1)}{2} +\frac{4}{3} \langle \sum_{i<j}P_{ij}^{s}\rangle
+ 2\langle \sum_{i<j}P_{ij}^{c}\rangle +4\langle \sum_{i<j}P_{ij}^{f}\rangle\nonumber\\
&  = n(n-10) +\frac{4}{3}S(S+1) +  2C_2[C_n] + 4C_2[F_n]
\end{align}

%For instance, $C_2$ for color singlet $=0$, 
%$C_2(I=3, (6,0)) = 18$, and $C_2(I=0, (0,3)) = 6$.
%The values of $C_2(SU(3)$ is tabulated in Table 1.

\section{Color matrix elements}\label{appB}

In this study, we have introduced a new set of color configurations, K(8). 
Solving the Schr\"odinger equation,  
we need to calculate the color matrix elements involving the new color configurations. 

For a color-singlet $(4q+1\bar q)$ system, 
we consider the matrix elements of the identity operator 1 (the overlap of the color wave functions) and
the $\Lambda_{ij}=\sum_a(\lambda^a_i\lambda^a_j)$ operator. Here $i=1-4$ labels the quarks ($4q$)
and $i=5$ is for the antiquark ($\bar q$). 
For the antiquark, we note the color operator is replaced by $\lambda^a\to-\lambda^{a*}$.

Several identity relations are useful.
\begin{eqnarray}
&& \sum_{i<j}\Lambda_{ij} 
= \frac{1}{2}\left( \sum_a(\sum_i \lambda^a_i)^2-\sum_i\sum_{a} (\lambda^a_i)^2 \right)\,,
\end{eqnarray}
and then
\begin{eqnarray}
&& \sum_{i<j\le 4}\Lambda_{ij}%= \frac{1}{2}\left( \sum_a (T^a(3))^2-4 \sum_{a} (T^a(3))^2 \right)
= \frac{1}{2}\left(\frac{16}{3} -4 \,\frac{16}{3}\right) = -8\,,
\end{eqnarray}
where we use $\sum_a (\lambda^a)^2 = 16/3$\,. 
A similar operator identity relation is obtained 
\begin{eqnarray}
&& \sum_{i=1}^4\Lambda_{i5} %= \frac{1}{2}\left( \sum_a(T^a_q+T^a_5)^2-\sum_{a} (T^a_q)^2 -\sum_{a} (T^a_5)^2\right)\nonumber\\
%= \frac{1}{2} \left(0-\frac{16}{3}-\frac{16}{3}\right) 
= -\frac{16}{3} \label{A3}\,.
\end{eqnarray}

The color octet $3q$ system is chosen in this study as $|(123)_{\bf 8}\rangle = |(12)_{\overline{\bf 3}} ,3;{\bf 8}\rangle$,
which results in
\begin{eqnarray}
&& \langle (123)_{\bf 8}|\Lambda_{12}| (123)_{\bf 8}\rangle =-\frac{16}{3}\\
&& \langle (123)_{\bf 8}|\Lambda_{13}| (123)_{\bf 8}\rangle =\langle (123)_{\bf 8}|\Lambda_{23}| (123)_{\bf 8}\rangle 
= \frac{1}{3}.
%&& \langle (12)3;8|\Lambda_{12}+\Lambda_{13}+\Lambda_{23}| (12)3;8\rangle = -2
\end{eqnarray}

\begin{widetext}
In order to calculate the color matrix elements,  
the color exchange operators for $ij$ quarks 
$P_{ij}= \frac{1}{2} \Lambda_{ij} + \frac{1}{3}$ are useful.
%\begin{eqnarray}
%&&P_{ij}= \frac{1}{2} \Lambda_{ij} + \frac{1}{3}.
%\end{eqnarray}
%because $\TT{i}{j}= \frac{1}{3}$ ($-\frac{2}{3}$) for the symmetric (antisymmetric) pair.
Here, we give some overlap matrix elements
\begin{eqnarray} 
&& \langle (123)_{\bf 1} (4\bar 5)_{\bf 1}| (124)_{\bf 1} (3\bar 5)_{\bf 1}\rangle 
%= \langle (123)_1 (4\bar 5)_1|P_{34}| (123)_1 (4\bar 5)_1\rangle 
= \frac{1}{2} \langle (123)_{\bf 1} (4\bar 5)_{\bf 1}|\Lambda_{34}| (123)_{\bf 1} (4\bar 5)_{\bf 1}\rangle+\frac{1}{3}
=\frac{1}{3}\\%\nonumber\\
&& \langle (123)_{\bf 8} (4\bar 5)_{\bf 8}| (124)_{\bf 8} (3\bar 5)_{\bf 8}\rangle 
%= \langle (123)_8 (4\bar 5)_8|P_{34}| (123)_8 (4\bar 5)_8\rangle 
= \frac{1}{2} \langle (123)_{\bf 8} (4\bar 5)_{\bf 8}|\Lambda_{34}| (123)_{\bf 8} (4\bar 5)_{\bf 8}\rangle +\frac{1}{3}
=-\frac{1}{3}\\%\nonumber\\
&& \langle (123)_{\bf 8} (4\bar 5)_{\bf 8}| (124)_{\bf 1} (3\bar 5)_{\bf 1}\rangle 
%= \langle (123)_8 (4\bar 5)_8|P_{34}| (123)_1 (4\bar 5)_1\rangle 
= \frac{1}{2} \langle (123)_{\bf 8} (4\bar 5)_{\bf 8}|\Lambda_{34}| (123)_{\bf 1} (4\bar 5)_{\bf 1}\rangle 
=\frac{2\sqrt{2}}{3}\\% \nonumber\\
&&  \langle (123)_{\bf 8} (4\bar 5)_{\bf 8}| (213)_{\bf 8} (4\bar 5)_{\bf 8}\rangle
=\frac{1}{2} \langle (123)_{\bf 8} (4\bar 5)_{\bf 8}|
\Lambda_{12}| (123)_{\bf 8} (4\bar 5)_{\bf 8}\rangle+\frac{1}{3} = -1\\%\nonumber\\
&&  \langle (123)_{\bf 8} (4\bar 5)_8| (132)_{\bf 8} (4\bar 5)_8\rangle
=\frac{1}{2} \langle (123)_{\bf 8} (4\bar 5)_{\bf 8}|
\Lambda_{23}| (123)_{\bf 8} (4\bar 5)_{\bf 8}\rangle+\frac{1}{3} = \frac{1}{2}\,.%\nonumber
\end{eqnarray}

Some of the matrix elements of the $\Lambda_{ij}$ operator used in the current study are 
\begin{eqnarray}
&& \langle (123)_{\bf 8} (4\bar 5)_{\bf 8}|\TT{1}{2}| (123)_{\bf 1} (4\bar 5)_{\bf 1}\rangle = 0 \\%\nonumber\\
&&  \langle (123)_{\bf 8} (4\bar 5)_{\bf 8}|\TT{1}{4}| (123)_{\bf 1} (4\bar 5)_{\bf 1}\rangle 
 =\langle (123)_{\bf 8} (4\bar 5)_{\bf 8}|\TT{2}{4}| (123)_{\bf 1} (4\bar 5)_{\bf 1}\rangle =- \frac{2\sqrt{2}}{3} \\%\nonumber\\
&& \langle (123)_{\bf 8} (4\bar 5)_{\bf 8}|\TT{3}{4}| (123)_{\bf 1} (4\bar 5)_{\bf 1}\rangle 
=\frac{4\sqrt{2}}{3}\\%\nonumber\\
&& \langle (123)_{\bf 8} (4\bar 5)_{\bf 8}|\TT{1}{5}| (123)_{\bf 1} (4\bar 5)_{\bf 1}\rangle 
 =\langle (123)_{\bf 8} (4\bar 5)_{\bf 8}|\TT{2}{5}| (123)_{\bf 1} (4\bar 5)_{\bf 1}\rangle = \frac{2\sqrt{2}}{3} \\%\nonumber\\
&& \langle (123)_{\bf 8} (4\bar 5)_{\bf 8}|\TT{3}{5}| (123)_{\bf 1} (4\bar 5)_{\bf 1}\rangle =-\frac{4\sqrt{2}}{3} \\%\nonumber\\
%&& \langle (123)_8 (4\bar 5)_8|\sum_{1\le i<j\le 3}\TT{i}{j}| (123)_8 (4\bar 5)_8\rangle 
% = -2\nonumber\\
&& \langle (123)_{\bf 8} (4\bar 5)_{\bf 8}|\TT{1}{4}| (123)_{\bf 8} (4\bar 5)_{\bf 8}\rangle 
 =\langle (123)_{\bf 8} (4\bar 5)_{\bf 8}|\TT{2}{4}| (123)_{\bf 8} (4\bar 5)_{\bf 8}\rangle = -\frac{7}{3}\\%\nonumber\\
&& \langle (123)_{\bf 8} (4\bar 5)_{\bf 8}|\TT{3}{4}| (123)_{\bf 8} (4\bar 5)_{\bf 8}\rangle =-\frac{4}{3}\\%\nonumber\\
%&& \langle (123)_8 (4\bar 5)_8|\sum_{1\le i<j\le 4}\TT{i}{j}| (123)_8 (4\bar 5)_8\rangle 
% = -8\nonumber\\
&& \langle (123)_{\bf 8} (4\bar 5)_{\bf 8}|\TT{1}{5}| (123)_{\bf 8} (4\bar 5)_{\bf 8}\rangle 
 =\langle (123)_{\bf 8} (4\bar 5)_{\bf 8}|\TT{2}{5}| (123)_{\bf 8} (4\bar 5)_{\bf 8}\rangle = -\frac{2}{3}\\%\nonumber\\
&& \langle (123)_{\bf 8} (4\bar 5)_{\bf 8}|\TT{3}{5}| (123)_{\bf 8} (4\bar 5)_{\bf 8}\rangle 
=-\frac{14}{3}\\%\nonumber\\
&& \langle (123)_{\bf 8} (4\bar 5)_{\bf 8}| \TT{4}{5}| (123)_{\bf 8} (4\bar 5)_{\bf 8}\rangle 
= \frac{2}{3}\,. %\nonumber
\end{eqnarray}
\end{widetext}
%% Create the reference section using BibTeX:
\bibliography{penta_25}

\end{document}